\documentclass[letterpaper,journal]{IEEEtran}
\usepackage[switch]{lineno}
\usepackage{amsmath,amsfonts,amssymb}
\usepackage{algorithmic}
\usepackage{algorithm}
\usepackage{array}
\usepackage[caption=false,font=normalsize,labelfont=sf,textfont=sf]{subfig}
\usepackage{textcomp}
\usepackage{stfloats}
\usepackage{url}
\usepackage{verbatim}
\usepackage{graphicx}
\usepackage{cite}
\usepackage[hidelinks, bookmarks=true,
bookmarksnumbered=true, bookmarkstype=toc,
colorlinks=true, citecolor=blue]{hyperref} 
\usepackage{color}

\begin{document}

\title{GLTCAM: Concept of Multi-color Millimeter and Submillimeter Camera for the Greenland Telescope}

\author{
    \IEEEauthorblockN{
        Shuhei Inoue,
        Tatsuya Takekoshi,
        Shinsuke Uno,
        Kazuki Watanabe,
        Taiki Sato,
        Toshihiro Tsuzuki,
        Satoru Mima,
        Tohru Taino,
        Kazuyuki Fujita,
        Shunichi Nakatsubo,
        Yuki Kimura,
        Chiko Otani,
        Ryohei Kawabe,
        Kotaro Kohno,
        and Tai Oshima}
    
\thanks{Manuscript received 25 September, 2025; revised 3 February, 2026; accepted 3 February. Date of publication 25 February.
{\it (Corresponding author: Shuhei Inoue.)}}
\thanks{Digital Object Identifier (DOI): 10.1109/TASC.2026.3664169}
\thanks{Shuhei Inoue and Kotaro Kohno are with the University of Tokyo, Tokyo 181-8588, Japan. (email: sinoue@ioa.s.u-tokyo.ac.jp)}
\thanks{Tatsuya Takekoshi is with Kitami Institute of Technology, Hokkaido 090-8507, Japan.}
\thanks{Shinsuke Uno and Chiko Otani are with RIKEN Center for Advanced Photonics, Saitama 351-0198, Japan.}
\thanks{Kazuki Watanabe, Ryohei Kawabe, and Tai Oshima are with National Astronomical Observatory of Japan (NAOJ), Tokyo 181-8588, Japan, 
and also with the Graduate University for Advanced Studies (SOKENDAI), Tokyo 181-8588, Japan.}
\thanks{Taiki Sato and Tohru Taino are with Saitama University, 255 Shimo-Okubo, Sakura-ku, Saitama 338-8570, Japan.}
\thanks{Toshihiro Tsuzuki is with National Astronomical Observatory of Japan (NAOJ), Tokyo 181-8588, Japan.}
\thanks{Satoru Mima is with National Institute of Information and Communications Technology, Kobe, Hyogo 651-2492, Japan.}
\thanks{Kazuyuki Fujita and Yuki Kimura are with Institute of Low Temperature Science, Hokkaido University, Sapporo 060-0819, Japan.}
\thanks{Shunichi Nakatsubo is with 
Institute of Space and Astronautical Science, Japan Aerospace Exploration Agency, Sagamihara 252-5210, Japan.}
}

\markboth{IEEE Transactions on Applied Superconductivity,~Vol.~36, No.~6, SEPTEMBER~2026}%
{Inoue \MakeLowercase{\textit{et al.}}: XXX Title}

\IEEEpubid{%
\raisebox{-3mm}{%
\makebox[\columnwidth]{%
\fbox{%
\parbox{\columnwidth}{\footnotesize
\copyright~2026 IEEE. Personal use of this material is permitted.
Permission from IEEE must be obtained for all other uses, in any current or
future media, including reprinting/republishing this material for advertising
or promotional purposes, creating new collective works, for resale or
redistribution to servers or lists, or reuse of any copyrighted component of
this work in other works.
}}}%
\hfill}
}


\maketitle

\begin{abstract}

To investigate the formation history of large-scale structure through the dynamics of galaxy clusters, we are developing a multi-color millimeter and submillimeter-wave continuum camera (GLTCAM) for deployment on the Greenland Telescope (GLT). GLTCAM will observe in six frequency bands—three in the millimeter range (150, 220, and 270~GHz) and three in the submillimeter range (350, 400, and 670~GHz). The optical design provides a compact configuration that fits within the GLT receiver cabin, while delivering diffraction-limited performance over an $18'$ field of view with minimal telecentricity error and distortion. A key advantage of this design is its uniform illumination footprint at the cold stop, which helps minimize thermal loading on both the detectors and the cryogenic stages. The focal plane module comprises a quasi-optical bandpass filter, a conical horn array coupled with planar ortho-mode transducers (OMTs), and a superconducting multi-color microwave kinetic inductance detector (MKID) array. Current development efforts are focused on the three-color millimeter-wave module. The detector array employs a single-layer coplanar waveguide (CPW) architecture, which simplifies fabrication and enables scalability to large-format arrays. GLTCAM aims for the early realization of next-generation wide-field, multi-color observations as a pathfinder for future large submillimeter telescopes.

\end{abstract}

\IEEEpubidadjcol

\vspace{5cm}

\section{Introduction}\label{sec1}
\IEEEpubidadjcol

\IEEEPARstart{G}{alaxy} clusters, the largest gravitationally bound objects in the universe, are thought to have evolved hierarchically through successive mergers over timescales comparable to the age of the universe \cite{2012Kravtsov}. Such mergers are expected to drive gas motions with velocities exceeding $\sim$1000 km/s \cite{2014Nelson}. Directly imaging these gas motions in individual clusters over a broad range of redshifts provides crucial insights into the formation history of large-scale cosmic structure.

\begin{figure}[t]
\centering
\includegraphics[width=1.0\linewidth]{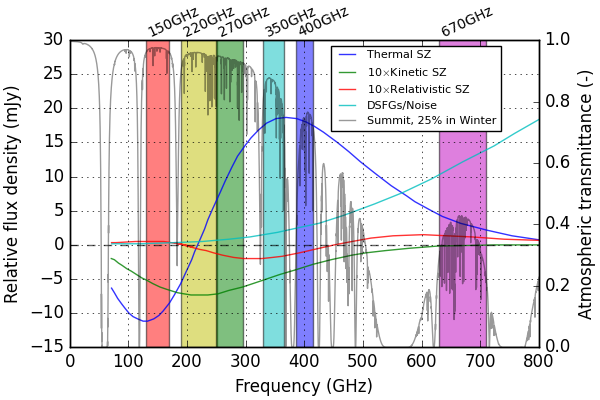}

\caption{Calculated spectral distortions due to the Sunyaev-Zel’dovich effects: 
\textbf{blue}, \textbf{green}, and \textbf{red} lines correspond to the thermal (tSZ), kinetic (kSZ), and relativistic (rSZ) components, respectively \cite{2018Mittal}. 
The \textbf{cyan} line shows contamination from dusty star-forming galaxies (DSFGs) \cite{2018Mittal}. 
The \textbf{black} line indicates the estimated atmospheric transmittance at the Greenland Summit Camp under 25\% opacity conditions in winter \cite{2016Matsushita}. The transmittance was calculated using the \textit{am} atmospheric model \cite{2024Paine}, assuming a zenith angle of 0$^\circ$.
\textbf{Shaded regions} denote the GLTCAM observation bands (150, 220, 270, 350, 400, and 670~GHz), chosen with reference to atmospheric transmission windows. 
Three bands in the millimeter (150, 220, and 270~GHz) and submillimeter (350, 400, and 670~GHz) ranges are observed simultaneously using dedicated focal plane modules.}

\label{fig:sz}
\end{figure}

\IEEEpubidadjcol

In recent years, the Sunyaev-Zel’dovich (SZ) effect has gained attention as a promising method for probing gas dynamics in galaxy clusters, owing to its redshift-independent surface brightness \cite{2014Kitayama,2019Mroczkowski}.
Figure \ref{fig:sz} presents an example of the spectral signature of the SZ effect, along with contamination from dusty star-forming galaxies (DSFGs) \cite{2018Mittal}. The kinetic SZ (kSZ) effect, which arises from the bulk motion of intracluster gas, is approximately an order of magnitude weaker ($\sim$ 1~mJy) than the thermal SZ (tSZ) effect. Therefore, to overcome the confusion limit imposed by unresolved astrophysical sources, 10-meter-class ground-based telescopes are required \cite{2011Bethermin}.

Accurate characterization of the SZ spectral shape and extraction of gas velocities requires observations across at least six frequency bands from millimeter to submillimeter wavelengths, including frequencies above 600~GHz where foreground dust emission dominates \cite{2012deBernardis}.

Recent efforts toward multi-color SZ observations—such as the two-band measurements performed with IRAM/NIKA \cite{2017Adam}—have made notable progress. However, velocity detection has thus far been limited to extremely fast-moving clusters ($\sim$3000~km/s), highlighting the need for broader bandwidths and more frequency bands.

In addition, a major challenge in ground-based observations is the effective removal of atmospheric emission fluctuations, which constitute the primary source of uncertainty. One effective approach involves applying principal component analysis (PCA) or similar techniques to identify and subtract correlated components across the field of view \cite{2018Rodriguez-Montoya}. However, when the field of view is smaller than the angular extent of the target cluster, such methods may inadvertently remove the cluster’s extended emission along with atmospheric signals. Given that the typical angular diameter of a galaxy cluster is $\sim 10'$ (e.g., a virial diameter of $\sim 2$~Mpc at $z = 0.2$), a field of view larger than this scale is required.

Based on these considerations, detecting cluster-scale motions through the kSZ effect necessitates millimeter and submillimeter-wave observations that simultaneously meet the following three key requirements: (1) high angular resolution using a telescope with an aperture exceeding 10~m, (2) broad bandwidth coverage with multi-color imaging across six or more frequency bands, and (3) a field of view wider than $10'$. Meeting these conditions would not only enable detailed studies of galaxy cluster dynamics but also facilitate a wide range of scientific applications, including submillimeter galaxy (SMG) surveys for probing cosmic star formation history, and rapid follow-up observations of transient astronomical events.

In this paper, we present an overview of the GLTCAM project, which aims to fulfill these three observational requirements.
Section~\ref{sec2} outlines the instrument specifications (\ref{sec2A}), the design of the wide-field optical system (\ref{sec2B}), and the development status of the focal plane module (\ref{sec2C}). Section~\ref{sec3} concludes the paper.

\section{GLTCAM}\label{sec2}

To realize wide-field, high-resolution, broadband, multi-color observations, we are developing a new millimeter- and submillimeter-wave multi-color\footnote{Throughout this paper, we use the term ``multi-color'' to describe the camera system capable of observing in multiple frequency bands, while the term ``multi-chroic'' refers specifically to detector technology that separates and detects multiple bands within a single pixel.} camera (GLTCAM). The instrument will be installed on the 12-meter Greenland Telescope (GLT) \cite{2023Chen}, which provides a relatively spacious cabin accommodating an $18'$ field of view and a surface accuracy of $16~\rm \mu$m, suitable for submillimeter observations.

The GLT enables observations of galaxy clusters in the northern sky, providing a complementary capability to wide-field SZ surveys conducted with southern-hemisphere telescopes such as SPT and ACT. In addition, synergistic studies are anticipated with galaxy cluster data obtained from X-ray satellites (e.g., XRISM) and optical facilities (e.g., Subaru).

\begin{table*}[t]
\caption{GLTCAM specification for observations, optics, and detectors}
\centering
\resizebox{\textwidth}{!}{
\begin{tabular}{c|c|cccccc}
\hline
& & & mm 3-color bands & & & submm 3-color bands& \\
 & & 150 GHz & 220 GHz & 270 GHz & 350 GHz & 400 GHz & 670 GHz\\ \hline
Observation & Frequency range (GHz) & 130--170  & 190--250 & 250--295 & 330--365 & 385--415 & 630--710 \\
& Resolution (arcmin, FWHM) & 42 & 27 & 23 & 18 & 16 & 9 \\ \hline
Optics & Telescope diameter (m) & \multicolumn{6}{c}{12} \\
 & Field of view (arcmin) & \multicolumn{6}{c}{18} \\
& F-number & \multicolumn{6}{c}{2.5}\\
& Components & \multicolumn{6}{c}{3 off-axis mirrors at 300~K (M3, M4, M5 in Fig. \ref{fig:optics_ray})}\\
& & \multicolumn{6}{c}{2 silicon lenses at 4~K (Lens \#1, \#2 in Fig. \ref{fig:optics_ray})}\\ \hline
Detectors & Optical coupling & \multicolumn{6}{c}{Conical-horn / OMT antenna-coupled} \\
& Detectors & \multicolumn{6}{c}{Microwave Kinetic Inductance Detectors (MKIDs)} \\
& Horn pitch (mm) & 5.5 & 5.5 & 5.5 & 2.2 & 2.2 & 2.2 \\
& Horn pitch ($F \lambda$) & 1.10 & 1.61 & 1.98 & 1.03 & 1.17 & 1.97 \\
& Pixels / $18'$ diameter & 637($=$91$\times$7) & 637 & 637 & 3829($=$547$\times$7) & 3829 & 3829 \\
\hline
\end{tabular}
}
\label{tab:table1}
\end{table*}

\subsection{Specification}\label{sec2A}

Table~\ref{tab:table1} summarizes the key specifications of GLTCAM.
The observing frequency bands are chosen to cover six bands between 130 and 710~GHz in order to capture thermal dust emission and to accurately characterize the spectral shape of the SZ effect. These bands are allocated within the available atmospheric transmission windows at the Greenland Summit Camp, with multiple bands assigned to broader windows \cite{2016Matsushita}.

A promising approach to multi-color imaging employs on-chip superconducting frequency filters integrated onto the detector wafer at the focal plane (e.g., \cite{2005Myers, 2013O'Brient, 2019Endo}).
This technique enables a compact optical system by allowing multiple bands to share a common focal plane and reduces inter-band systematic errors. In this project, we adopt this approach and employ an on-chip multi-color microwave kinetic inductance detector (MKID) array.

For optical coupling of the focal plane detectors, we employ a conical horn antenna combined with a planar ortho-mode transducer (OMT), as implemented in CMB experiments such as CMB-S4 \cite{2017Abitbol} (see Section~\ref{sec2C} for details). In this configuration, the usable bandwidth of the conical horn antenna is typically limited to $\sim 2.3$. Moreover, to maintain mapping speed, the horn pitch must remain within approximately $1F\lambda$--$2F\lambda$ ($F$ denotes the F-number and $\lambda$ the wavelength), beyond which performance degrades significantly \cite{2002Griffin, 2002Halverson, 2010Padin}. Consequently, the practical fractional bandwidth of a single detector array is limited to $\sim 2.3$, making it infeasible to cover all six bands with a single array.

To maximize the number of simultaneously observable bands under these constraints, we plan to develop two focal plane modules: a three-band millimeter-wave module (150, 220, and 270~GHz; fractional bandwidth 2.27:1) and a three-band submillimeter-wave module (350, 400, 670~GHz; 2.15:1). The millimeter-wave module will be developed and deployed on the GLT first, followed by replacement with the submillimeter-wave module. This sequential approach will provide full coverage of all six bands. Each focal plane module will be populated to cover the full $18'$ field of view, using the horn pitch values listed in Table~\ref{tab:table1}.

\subsection{Optics}\label{sec2B}

\begin{figure}[t]
\centering
\includegraphics[width=0.8\linewidth]{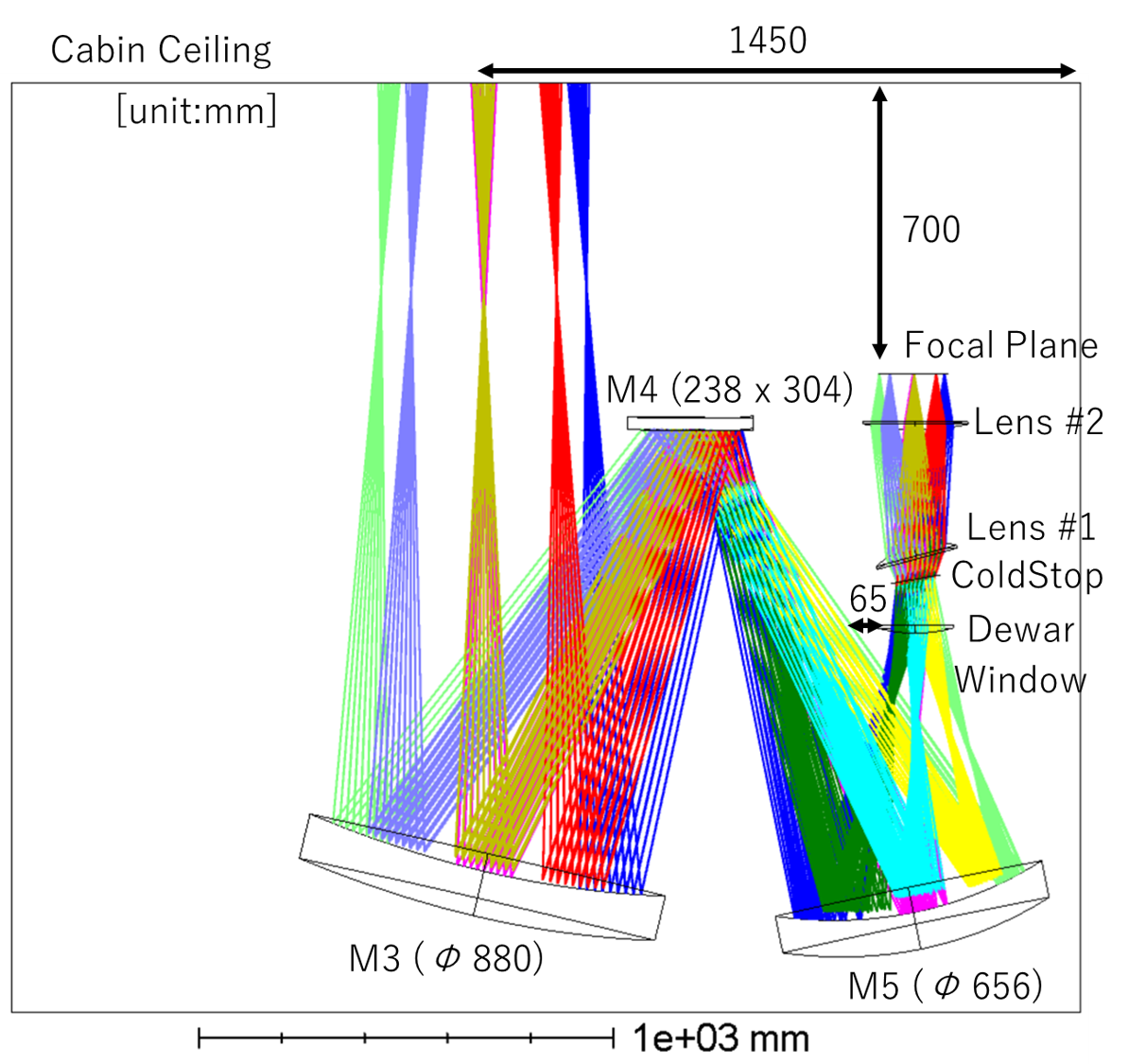}
\caption{Schematic diagram of the GLTCAM optical system. Radio signals collected by the GLT’s Cassegrain optics are relayed by three 300~K mirrors (M3--M5) into the cryostat.
Inside the cryostat, through a cold stop and two silicon lenses at 4~K stage, the beam is focused onto the focal plane module at the 150~mK stage.
}
\label{fig:optics_ray}
\end{figure}

\begin{figure*}[t]
\centering
\includegraphics[width=1.0\linewidth]{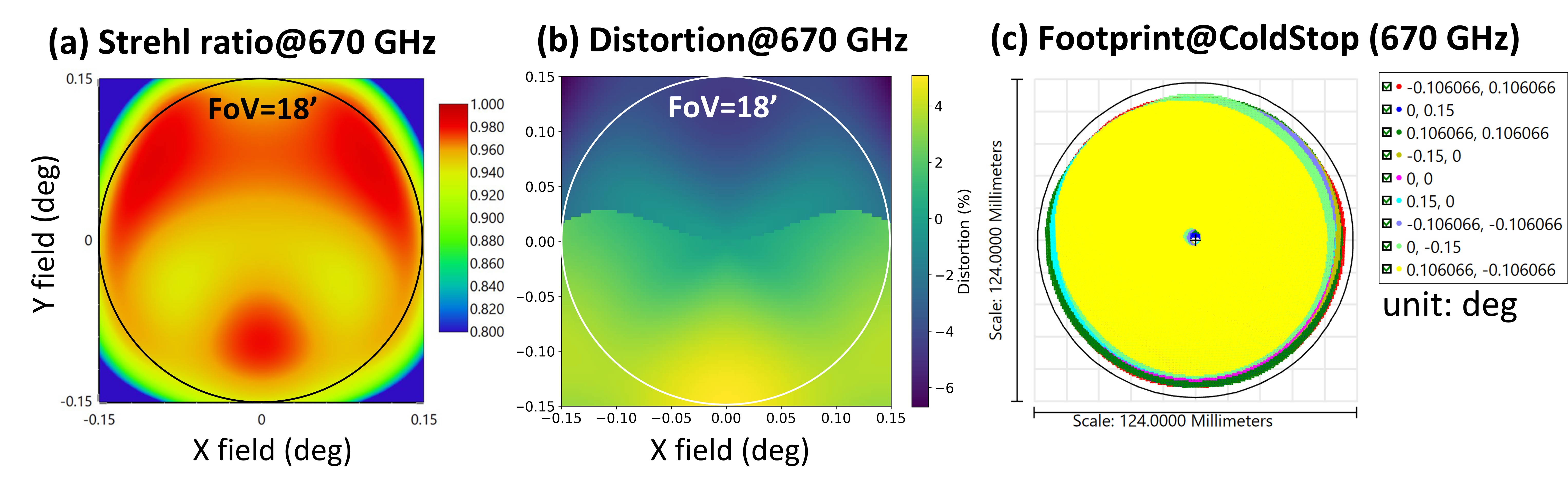}
\caption{Simulated optical performance of GLTCAM at 670~GHz by Ansys Zemax OpticStudio. (a) Strehl ratio across the focal plane, showing values above $0.93$ over the full $18'$ field of view. (b) Distortion map at the focal plane, with distortion suppressed below $5.1$\%. (c) Footprint of the beam at the cold stop, demonstrating uniform illumination across the field.}
\label{fig:optics_results_2}
\end{figure*}

Figure~\ref{fig:optics_ray} shows the signal path and configuration of the wide-field optical design for GLTCAM. The optics were designed to be compact enough to fit within the GLT receiver cabin, while achieving an $18'$ field of view with high Strehl ratio, low aberration, and high telecentricity. To this end, key parameters such as the mirror arrangement, the geometry of the vacuum window, and the tilt angles of the cold stop and silicon lenses were optimized as described below.

The incoming radio signal, matched to the GLT’s Cassegrain optics ($F$/8) and cabin, is relayed by three free-form surface mirrors (M3, M4, M5) in the 300K optics and directed into the cryostat. At the interface between the 300K optics and the cryostat, we employ a vacuum window made of ultra-high-molecular-weight polyethylene (UHMWPE; refractive index $n=1.52$, thickness 8~mm) with a cylindrical surface on one side.
The anti-reflection coating is realized with a five-layer stack of PTFE and porous PTFE, providing an ultra-broadband average reflectance of 0.2\% over the 130--710~GHz range \cite{2026Naganuma}.

Inside the cryostat, infrared-blocking filters mounted at the apertures of the 50~K and 4~K radiation shields suppress 300~K thermal radiation.
The incoming beam is collimated by a tilted cold stop ($\theta = 10^\circ$) at the entrance of the 4~K stage, and then focused by two plano-convex silicon lenses ($n = 3.42$): Lens~\#1 ($15.3$~mm thick, $\theta = 15^\circ$) and Lens~\#2 ($14.8$~mm thick, $\theta = 0^\circ$), onto the focal plane located at the 150~mK stage. Cooling is provided by a pulse tube refrigerator (SHI SRP-082B2S) down to 4~K, with a Chase continuous mini-dilutor for further cooling to 150~mK, thereby maintaining the detectors continuously at base temperature.

This optical design achieves a compact mirror layout that fits within the GLT receiver cabin, while accommodating the cryogenic optics in a compact $\phi 25$~mm cylindrical volume. Even at the highest frequency band of 670~GHz, the system delivers diffraction-limited performance across the $18'$ field of view, achieving a Strehl ratio above $0.93$ (Fig.~\ref{fig:optics_results_2}(a)). Furthermore, distortion is suppressed to below $5.1$\% (Fig.~\ref{fig:optics_results_2}(b)) and the telecentric error is limited to $0.33^\circ$. Moreover, owing to the uniform footprint at the cold stop across the field (Fig.~\ref{fig:optics_results_2}(c)), the design reduces thermal loading on both the detectors and the sub-Kelvin stages.

\subsection{Focal Plane Module}\label{sec2C}

\begin{figure*}[t]
\centering
\includegraphics[width=1.0\linewidth]{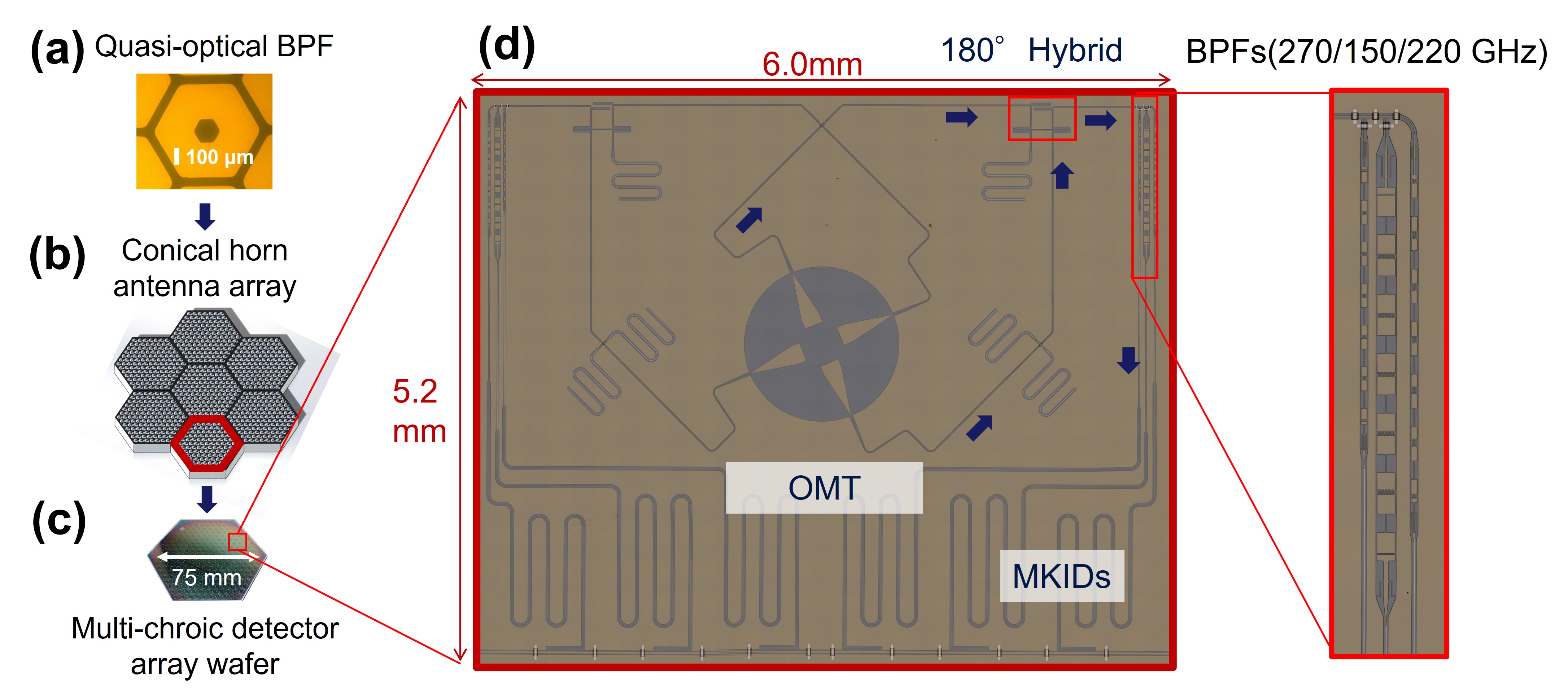}
\caption{Focal plane module of GLTCAM. 
Seven hexagonal arrays (each $\sim 75$~mm in diameter) are arranged to cover the $18'$ field of view in both the millimeter-wave and submillimeter-wave modules, consisting of 91 pixels and 547 pixels per array, respectively.
(a) Micrograph of a quasi-optical bandpass filter \cite{2020Uno}.
(b) Schematic of the conical horn array. 
(c) Schematic of the superconducting multi-chroic detector array. 
(d) Micrograph of a prototype superconducting multi-chroic detector for millimeter-wave bands: horn-coupled OMT, 180$^\circ$ hybrid coupler, three-color bandpass filters (150, 220, and 270~GHz), and MKIDs.}

\label{fig:focalplane_array_2}
\end{figure*}

This section describes the signal flow and configuration of the focal plane module, which consists of a quasi-optical filter, a conical horn antenna array, and a superconducting detector array. Figure~\ref{fig:focalplane_array_2} presents a conceptual illustration of the focal plane module used in GLTCAM.
The quasi-optical bandpass filter is positioned directly in front of the conical horn array (Figs.~\ref{fig:focalplane_array_2}(a) and (b)) and selectively transmits signals in either the millimeter-wave range (130--295~GHz) or the submillimeter-wave range (330--710~GHz). The incoming free-space signal then passes through the horn and is coupled to an OMT on the superconducting detector chip (Figs.~\ref{fig:focalplane_array_2}(c) and (d)), where it is converted into an on-chip millimeter-/submillimeter-wave circuit that simultaneously acquires both polarizations.  
The fundamental TE$_{11}$ mode of the circular waveguide is combined at the difference port of a 180$^\circ$ hybrid coupler, while unwanted even modes are terminated at the sum port. The signal is subsequently spectrally separated by on-chip bandpass filters into three bands---150, 220, and 270~GHz for the millimeter-wave module, or 350, 400, and 670~GHz for the submillimeter-wave module---with each frequency band directed to and detected by a dedicated MKID. Thus, there are six MKIDs in the pixel design shown in Fig. \ref{fig:focalplane_array_2}(d) for each frequency band and linear polarization.

The quasi-optical bandpass filter is based on a three-layer metal-mesh structure fabricated using flexible printed circuit (FPC) technology, which has demonstrated broadband transmission across 170--520~GHz \cite{2020Uno}. For GLTCAM, the filter design is scaled to match the frequency bands of the millimeter-wave and submillimeter-wave modules.  
The horn array is fabricated from a silicon--aluminum composite, whose thermal expansion coefficient is about one third that of standard aluminum. Although the machinability of this composite is inferior to that of aluminum alloys, prototype horn arrays have already been successfully manufactured and their beam patterns evaluated \cite{2022Takekoshi}. In GLTCAM, the array will be scaled to accommodate the required number of horns to tile the field of view.  
The superconducting detector array employs high resistivity silicon as the dielectric substrate (relative permittivity $\epsilon_{r} = 11.45$; thickness $200~\mu$m) and either niobium (Nb) or niobium titanium nitride (NbTiN) as the superconducting material. The baseline on-chip RF circuitry adopts a coplanar waveguide (CPW) structure, enabling a mostly single-layer implementation and simplifying fabrication, except for localized multilayer features such as the OMT, crossovers, and bridges.  

The OMT incorporates probes fabricated on a silicon nitride (SiN$_x$) membrane together with a backshort, in addition to silicon, and has achieved a wideband design solution with a fractional bandwidth of 2.28:1 \cite{Uno2024_PhD}.  
The 180$^\circ$ hybrid coupler is likewise realized in a CPW-based single-layer configuration, and a scale model operating in the 6--14~GHz range has demonstrated the required 2.3:1 fractional bandwidth \cite{2024Inoue}; millimeter- and submillimeter-wave implementations are planned for future development.  
The on-chip bandpass filter employs an optimized triplexer architecture that splits the incoming signal into three frequency paths. Each path is followed by an individual band-defining filter, realized as an eight-stage Chebyshev filter composed of cascaded planar inductors and capacitors. The triplexer's transmittance has been simulated, and a $\sim 20 \times$ scale model of the 150~GHz band filter has been fabricated and characterized \cite{2025Uno}. 
The MKIDs adopt a hybrid structure composed of aluminum and either Nb or NbTiN, which is considered effective for achieving high sensitivity \cite{2011Yates, 2013Janssen}.

The MKID readout system adopts a multiplexed architecture within a 2~GHz bandwidth \cite{2016Rantwijk}, as successfully demonstrated in field observations with DESHIMA \cite{2019Endo}. A future transition to a scheme based on a Radio Frequency System-on-Chip (RFSoC) \cite{2024Smith} is also planned, enabling further improvements in compactness, power efficiency, and readout speed.

\section{Conclusion}\label{sec3}
We have presented an overview of the GLTCAM project, which aims to realize wide-field, high-resolution, ultra-broadband six-band imaging in the millimeter and submillimeter ranges to advance the study of galaxy cluster dynamics. The optical design is sufficiently compact to fit within the GLT receiver cabin, while achieving diffraction-limited performance with a high Strehl ratio across the full field, together with low distortion and telecentric error. In addition, the uniform footprint at the cold stop reduces thermal loading on both the detectors and the sub-Kelvin stages. The focal plane module integrates quasi-optical filters, conical horn antenna arrays, and superconducting multi-chroic MKID arrays. At present, the design, fabrication, and performance evaluation of the three-band millimeter-wave module are in progress. To enable large-scale array integration, the superconducting multi-chroic detectors are being developed using a single-layer CPW-based structure that minimizes fabrication complexity. These wide-field camera development efforts not only advance submillimeter astronomy, but also establish foundational technologies for future wide-field instruments on next-generation large submillimeter telescopes such as the Large Submillimeter Telescope (LST) \cite{2016Kawabe} and the Atacama Large Aperture Submillimeter Telescope (AtLAST) \cite{2025Mroczkowski}.

\section*{Acknowledgements}
We thank the anonymous reviewers for their careful review and valuable comments.
This study was carried out in cooperation with the Advanced Technology Center of the National Astronomical Observatory of Japan (NAOJ).
This study was carried out under the Joint Research Program of the Institute of Low Temperature Science, Hokkaido University (25G014, 24G016, 23G012, 22G007, 21G006, and 20G013).
This work was supported by the Joint Development Research, the NAOJ Research Coordination Committee, NINS (NAOJ-RCC-2201-0101 and NAOJ-RCC-2101-0101).
S.~Inoue was supported by FoPM, WINGS Program, the University of Tokyo.
T.~Takekoshi was supported by the MEXT Leading Initiative for Excellent Young Researchers (Grant No. JPMXS0320200188).
K.~Watanabe is supported by JST SPRING, Japan Grant Number JPMJSP2104.
This work was supported in part by RIKEN Special Postdoctoral Researcher Program,
JSPS KAKENHI Grant Numbers JP25KJ0981, JP24K22911, JP24H00004, JP23K25905, JP23K25879, JP23K20035, and JP23H00121, 
the Murata Science and Education Foundation, the Nakajima Foundation, and the Sumitomo Foundation (Basic Science Research Grant No. 2200541).

\bibliographystyle{IEEEtran}
\bibliography{references}

@article{2002Griffin,
    author = "Griffin, Matthew J. and Bock, James J. and Gear, Walter K.",
    title = "{{The Relative performance of filled and feedhorn-coupled focal-plane architectures}}",
    eprint = "astro-ph/0205264",
    archivePrefix = "arXiv",
    doi = "10.1364/AO.41.006543",
    journal = "Appl. Opt.",
    volume = "41",
    pages = "6543",
    year = "2002"
}

@misc{2002Halverson,
  author       = {Nils Halverson},
  title        = {{A Sunyaev-Zel’dovich Effect Survey with the APEX Telescope}},
  howpublished = {\url{https://online.kitp.ucsb.edu/online/cmb_c02/minisession1/pdf0/Halverson2.pdf}},
  year         = {2002},
  note         = {Accessed: Aug. 08, 2025}
}

@article{2005Myers,
    author = {Myers, Michael J. and Holzapfel, William and Lee, Adrian T. and O’Brient, Roger and Richards, P. L. and Tran, Huan T. and Ade, Peter and Engargiola, Greg and Smith, Andy and Spieler, Helmuth},
    title = {{An antenna-coupled bolometer with an integrated microstrip bandpass filter}},
    journal = {Applied Physics Letters},
    volume = {86},
    number = {11},
    pages = {114103},
    year = {2005},
    month = {03},
    issn = {0003-6951},
    doi = {10.1063/1.1879115},
    url = {https://doi.org/10.1063/1.1879115},
    eprint = {https://pubs.aip.org/aip/apl/article-pdf/doi/10.1063/1.1879115/14641567/114103\_1\_online.pdf},
}

@article{2010Padin,
author = {Stephen Padin},
journal = {Appl. Opt.},
number = {3},
pages = {479--483},
publisher = {Optica Publishing Group},
title = {{Mapping speed for an array of corrugated horns}},
volume = {49},
month = {Jan},
year = {2010},
url = {https://opg.optica.org/ao/abstract.cfm?URI=ao-49-3-479},
doi = {10.1364/AO.49.000479},
}

@article{2011Bethermin,
	author = {{Béthermin, M.} and {Dole, H.} and {Lagache, G.} and {Le Borgne, D.} and {Penin, A.}},
	title = {{Modeling the evolution of infrared galaxies: a parametric  backward evolution model}},
	DOI= "10.1051/0004-6361/201015841",
	url= "https://doi.org/10.1051/0004-6361/201015841",
	journal = {A\&A},
	year = 2011,
	volume = 529,
	pages = "A4",
	month = "",
}

@article{2011Yates,
    author = {Yates, S. J. C. and Baselmans, J. J. A. and Endo, A. and Janssen, R. M. J. and Ferrari, L. and Diener, P. and Baryshev, A. M.},
    title = {{Photon noise limited radiation detection with lens-antenna coupled microwave kinetic inductance detectors}},
    journal = {Applied Physics Letters},
    volume = {99},
    number = {7},
    pages = {073505},
    year = {2011},
    month = {08},
    issn = {0003-6951},
    doi = {10.1063/1.3624846},
    url = {https://doi.org/10.1063/1.3624846},
    eprint = {https://pubs.aip.org/aip/apl/article-pdf/doi/10.1063/1.3624846/14455756/073505\_1\_online.pdf},
}

@article{2012deBernardis,
	author = {{de Bernardis, P.} and {Colafrancesco, S.} and {D’Alessandro, G.} and {Lamagna, L.} and {Marchegiani, P.} and {Masi, S.} and {Schillaci, A.}},
	title = {{Low-resolution spectroscopy of the Sunyaev-Zel’dovich effect
          and estimates of cluster parameters}},
	DOI= "10.1051/0004-6361/201118062",
	url= "https://doi.org/10.1051/0004-6361/201118062",
	journal = {A\&A},
	year = 2012,
	volume = 538,
	pages = "A86",
	month = "",
}

@article{2012Kravtsov,
   author = "Kravtsov, Andrey V. and Borgani, Stefano",
   title = {{{Formation of Galaxy Clusters}}}, 
   journal= "Annual Review of Astronomy and Astrophysics",
   year = "2012",
   volume = "50",
   number = "Volume 50, 2012",
   pages = "353-409",
   doi = "https://doi.org/10.1146/annurev-astro-081811-125502",
   url = "https://www.annualreviews.org/content/journals/10.1146/annurev-astro-081811-125502",
   publisher = "Annual Reviews",
   issn = "1545-4282",
   type = "Journal Article",
  }

@article{2013Janssen,
    author = {Janssen, R. M. J. and Baselmans, J. J. A. and Endo, A. and Ferrari, L. and Yates, S. J. C. and Baryshev, A. M. and Klapwijk, T. M.},
    title = {{High optical efficiency and photon noise limited sensitivity of microwave kinetic inductance detectors using phase readout}},
    journal = {Applied Physics Letters},
    volume = {103},
    number = {20},
    pages = {203503},
    year = {2013},
    month = {11},
    issn = {0003-6951},
    doi = {10.1063/1.4829657},
    url = {https://doi.org/10.1063/1.4829657},
    eprint = {https://pubs.aip.org/aip/apl/article-pdf/doi/10.1063/1.4829657/14286378/203503\_1\_online.pdf},
}

@article{2014Kitayama,
    author = {Kitayama, Tetsu},
    title = {{Cosmological and astrophysical implications of the Sunyaev–Zel’dovich effect}},
    journal = {Progress of Theoretical and Experimental Physics},
    volume = {2014},
    number = {6},
    pages = {06B111},
    year = {2014},
    month = {06},
    issn = {2050-3911},
    doi = {10.1093/ptep/ptu055},
    url = {https://doi.org/10.1093/ptep/ptu055},
    eprint = {https://academic.oup.com/ptep/article-pdf/2014/6/06B111/4444647/ptu055.pdf},
}

@article{2014Nelson,
doi = {10.1088/0004-637X/792/1/25},
url = {https://dx.doi.org/10.1088/0004-637X/792/1/25},
year = {2014},
month = {aug},
publisher = {The American Astronomical Society},
volume = {792},
number = {1},
pages = {25},
author = {Nelson, Kaylea and Lau, Erwin T. and Nagai, Daisuke},
title = {{HYDRODYNAMIC SIMULATION OF NON-THERMAL PRESSURE PROFILES OF GALAXY CLUSTERS}},
journal = {The Astrophysical Journal}
}

@inproceedings{2016Kawabe,
author = {Ryohei Kawabe and Kotaro Kohno and Yoichi Tamura and Tatsuya Takekoshi and Tai Oshima and Shun Ishii},
title = {{New 50-m-class single-dish telescope: Large Submillimeter Telescope (LST)}},
volume = {9906},
booktitle = {Ground-based and Airborne Telescopes VI},
editor = {Helen J. Hall and Roberto Gilmozzi and Heather K. Marshall},
organization = {International Society for Optics and Photonics},
publisher = {SPIE},
pages = {990626},
year = {2016},
doi = {10.1117/12.2232202},
URL = {https://doi.org/10.1117/12.2232202}
}

@ARTICLE{2016Rantwijk,
  author={van Rantwijk, Joris and Grim, Martin and van Loon, Dennis and Yates, Stephen and Baryshev, Andrey and Baselmans, Jochem},
  journal={IEEE Transactions on Microwave Theory and Techniques}, 
  title={{Multiplexed Readout for 1000-Pixel Arrays of Microwave Kinetic Inductance Detectors}}, 
  year={2016},
  volume={64},
  number={6},
  pages={1876-1883},
  doi={10.1109/TMTT.2016.2544303}}

@article{2017Adam,
	author = {{Adam, R.} and {Bartalucci, I.} and {Pratt, G. W.} and {Ade, P.} and {André, P.} and {Arnaud, M.} and {Beelen, A.} and {Benoît, A.} and {Bideaud, A.} and {Billot, N.} and {Bourdin, H.} and {Bourrion, O.} and {Calvo, M.} and {Catalano, A.} and {Coiffard, G.} and {Comis, B.} and {D’Addabbo, A.} and {De Petris, M.} and {Démoclès, J.} and {Désert, F.-X.} and {Doyle, S.} and {Egami, E.} and {Ferrari, C.} and {Goupy, J.} and {Kramer, C.} and {Lagache, G.} and {Leclercq, S.} and {Macías-Pérez, J.-F.} and {Maurogordato, S.} and {Mauskopf, P.} and {Mayet, F.} and {Monfardini, A.} and {Mroczkowski, T.} and {Pajot, F.} and {Pascale, E.} and {Perotto, L.} and {Pisano, G.} and {Pointecouteau, E.} and {Ponthieu, N.} and {Revéret, V.} and {Ritacco, A.} and {Rodriguez, L.} and {Romero, C.} and {Ruppin, F.} and {Schuster, K.} and {Sievers, A.} and {Triqueneaux, S.} and {Tucker, C.} and {Zemcov, M.} and {Zylka, R.}},
	title = {{Mapping the kinetic Sunyaev-Zel’dovich effect toward  MACS J0717.5+3745 with NIKA }},
	DOI= "10.1051/0004-6361/201629182",
	url= "https://doi.org/10.1051/0004-6361/201629182",
	journal = {A\&A},
	year = 2017,
	volume = 598,
	pages = "A115",
}

@article{2016Matsushita,
doi = {10.1088/1538-3873/129/972/025001},
url = {https://dx.doi.org/10.1088/1538-3873/129/972/025001},
year = {2016},
month = {dec},
publisher = {The Astronomical Society of the Pacific},
volume = {129},
number = {972},
pages = {025001},
author = {Matsushita, Satoki and Asada, Keiichi and Martin-Cocher, Pierre L. and Chen, Ming-Tang and Ho, Paul T. P. and Inoue, Makoto and Koch, Patrick M. and Paine, Scott N. and Turner, David D.},
title = {{3.5 Year Monitoring of 225 GHz Opacity at the Summit of Greenland}},
journal = {Publications of the Astronomical Society of the Pacific}
}

@techreport{2017Abitbol,
  author       = {Abitbol, Maximilian H. and Zeeshan, Ahmed and Darcy, Barron and Basu Thakur, Ritoban and Bender, Amy N. and Benson, Bradford A. and Bischoff, Colin A. and Bryan, Sean A. and Carlstrom, John E. and Chang, Clarence L. and others},
  title        = {{CMB-S4 Technology Book, First Edition}},
  institution  = {Michigan U.; Lawrence Berkeley National Laboratory (LBNL), Berkeley, CA (United States); Cornell U.; UC, Berkeley; Brown U.; NASA, Goddard; Minnesota U.; Villanova U.; McGill U.; Wisconsin U., Madison; Stanford U.; Princeton U.; NIST, Boulder; Illinois U., Urbana (main); Arizona State U.; Colorado U.; Harvard U.; Columbia U.; Fermi National Accelerator Laboratory (FNAL), Batavia, IL (United States); Caltech; Case Western Reserve U.; Caltech, JPL; Argonne National Laboratory (ANL), Argonne, IL (United States); Cincinnati U.; APC, Paris; Stockholm U.; SLAC National Accelerator Laboratory (SLAC), Menlo Park, CA (United States); Chicago U.},
  doi          = {10.2172/1414402},
  url          = {https://www.osti.gov/biblio/1414402},
  place        = {United States},
  year         = {2017},
  month        = {06}}

@article{2018Mittal,
doi = {10.1088/1475-7516/2018/02/032},
url = {https://dx.doi.org/10.1088/1475-7516/2018/02/032},
year = {2018},
month = {feb},
volume = {2018},
number = {02},
pages = {032},
author = {Mittal, Avirukt and de Bernardis, Francesco and Niemack, Michael D.},
title = {{Optimizing measurements of cluster velocities and temperatures for CCAT-prime and future surveys}},
journal = {Journal of Cosmology and Astroparticle Physics}
}

@article{2018Rodriguez-Montoya,
doi = {10.3847/1538-4365/aaa83c},
url = {https://dx.doi.org/10.3847/1538-4365/aaa83c},
year = {2018},
month = {mar},
publisher = {The American Astronomical Society},
volume = {235},
number = {1},
pages = {12},
author = {Rodríguez-Montoya, Iván and Sánchez-Argüelles, David and Aretxaga, Itziar and Bertone, Emanuele and Chávez-Dagostino, Miguel and Hughes, David H. and Montaña, Alfredo and Wilson, Grant W. and Zeballos, Milagros},
title = {{Multiple-component Decomposition from Millimeter Single-channel Data}},
journal = {The Astrophysical Journal Supplement Series}
}

@Article{2019Endo,
author={Endo, Akira
and Karatsu, Kenichi
and Tamura, Yoichi
and Oshima, Tai
and Taniguchi, Akio
and Takekoshi, Tatsuya
and Asayama, Shin'ichiro
and Bakx, Tom J. L. C.
and Bosma, Sjoerd
and Bueno, Juan
and Chin, Kah Wuy
and Fujii, Yasunori
and Fujita, Kazuyuki
and Huiting, Robert
and Ikarashi, Soh
and Ishida, Tsuyoshi
and Ishii, Shun
and Kawabe, Ryohei
and Klapwijk, Teun M.
and Kohno, Kotaro
and Kouchi, Akira
and Llombart, Nuria
and Maekawa, Jun
and Murugesan, Vignesh
and Nakatsubo, Shunichi
and Naruse, Masato
and Ohtawara, Kazushige
and Pascual Laguna, Alejandro
and Suzuki, Junya
and Suzuki, Koyo
and Thoen, David J.
and Tsukagoshi, Takashi
and Ueda, Tetsutaro
and de Visser, Pieter J.
and van der Werf, Paul P.
and Yates, Stephen J. C.
and Yoshimura, Yuki
and Yurduseven, Ozan
and Baselmans, Jochem J. A.},
title={{First light demonstration of the integrated superconducting spectrometer}},
journal={Nature Astronomy},
year={2019},
month={Nov},
day={01},
volume={3},
number={11},
pages={989-996},
issn={2397-3366},
doi={10.1038/s41550-019-0850-8},
url={https://doi.org/10.1038/s41550-019-0850-8}
}

@Article{2019Mroczkowski,
author={Mroczkowski, Tony
and Nagai, Daisuke
and Basu, Kaustuv
and Chluba, Jens
and Sayers, Jack
and Adam, R{\'e}mi
and Churazov, Eugene
and Crites, Abigail
and Di Mascolo, Luca
and Eckert, Dominique
and Macias-Perez, Juan
and Mayet, Fr{\'e}d{\'e}ric
and Perotto, Laurence
and Pointecouteau, Etienne
and Romero, Charles
and Ruppin, Florian
and Scannapieco, Evan
and ZuHone, John},
title={{Astrophysics with the Spatially and Spectrally Resolved Sunyaev-Zeldovich Effects}},
journal={Space Science Reviews},
year={2019},
month={Feb},
day={12},
volume={215},
number={1},
pages={17},
issn={1572-9672},
doi={10.1007/s11214-019-0581-2},
url={https://doi.org/10.1007/s11214-019-0581-2}
}

@article{2020Uno,
author = {Shinsuke Uno and Tatsuya Takekoshi and Tai Oshima and Keisuke Yoshioka and Kah Wuy Chin and Kotaro Kohno},
journal = {Appl. Opt.},
number = {13},
pages = {4143--4150},
publisher = {Optica Publishing Group},
title = {{Demonstration of wideband metal mesh filters for submillimeter astrophysics using flexible printed circuits}},
volume = {59},
month = {May},
year = {2020},
url = {https://opg.optica.org/ao/abstract.cfm?URI=ao-59-13-4143},
doi = {10.1364/AO.389605},
}

@Article{2022Takekoshi,
author={Takekoshi, Tatsuya
and Lee, Kianhong
and Chin, Kah Wuy
and Uno, Shinsuke
and Naganuma, Toyo
and Inoue, Shuhei
and Niwa, Yuka
and Fujita, Kazuyuki
and Kouchi, Akira
and Nakatsubo, Shunichi
and Mima, Satoru
and Oshima, Tai},
title={{Material Properties of a Low Contraction and Resistivity Silicon--Aluminum Composite for Cryogenic Detectors}},
journal={Journal of Low Temperature Physics},
year={2022},
month={Dec},
day={01},
volume={209},
number={5},
pages={1143-1150},
issn={1573-7357},
doi={10.1007/s10909-022-02795-9},
url={https://doi.org/10.1007/s10909-022-02795-9}
}

@article{2023Chen,
doi = {10.1088/1538-3873/acf072},
url = {https://dx.doi.org/10.1088/1538-3873/acf072},
year = {2023},
month = {sep},
publisher = {The Astronomical Society of the Pacific},
volume = {135},
number = {1051},
pages = {095001},
author = {Chen, Ming-Tang and Asada, Keiichi and Matsushita, Satoki and Raffin, Philippe and Inoue, Makoto and Ho, Paul T. P. and Han, Chih-Chiang and Kubo, Derek and Norton, Timothy and Patel, Nimesh A. and Nystrom, George and Huang, Chih-Wei L. and Martin-Cocher, Pierre and Yi Koay, Jun and Romero-Cañizales, Cristina and Liu, Ching-Tang and Huang, Teddy and Liu, Kuan-Yu and Wei, Tashun and Chang, Shu-Hao and Chilson, Ryan and Oshiro, Peter and Jiang, Homin and Li, Chao-Te and Bower, Geoffrey and Shaw, Paul and Nishioka, Hiroaki and Koch, Patrick M. and Chen, Chung-Cheng and Srinivasan, Ranjani and Rao, Ramprasad and Snow, William and Jinchi, Hao and Han, Kuo-Chang and Chang, Song-Chu and Lu, Li-Ming and Ogawa, Hideo and Kimura, Kimihiro and Hasegawa, Yutaka and Pu, Hung-Yi and Koyama, Shoko and Nakamura, Masanori and Bintley, Daniel and Walther, Craig and Friberg, Per and Dempsey, Jessica and Sriharan, T. K. and Srikanth, Sivasankaran and Doeleman, Sheperd S. and Brissenden, Roger and Algaba Marcos, Juan-Carlos and Jeter, Britt and Kuo, Cheng-Yu and Park, Jongho},
title = {{The Greenland Telescope—Construction, Commissioning, and Operations in Pituffik}},
journal = {Publications of the Astronomical Society of the Pacific}
}

@article{2024Smith,
    author = {Smith, Jennifer Pearl and Bailey, John I., III and Cuda, Aled and Zobrist, Nicholas and Mazin, Benjamin A.},
    title = {{MKIDGen3: Energy-resolving, single-photon-counting microwave kinetic inductance detector readout on a radio frequency system-on-chip}},
    journal = {Review of Scientific Instruments},
    volume = {95},
    number = {11},
    pages = {114705},
    year = {2024},
    month = {11},
    issn = {0034-6748},
    doi = {10.1063/5.0225768},
    url = {https://doi.org/10.1063/5.0225768},
    eprint = {https://pubs.aip.org/aip/rsi/article-pdf/doi/10.1063/5.0225768/20241338/114705\_1\_5.0225768.pdf},
}

@phdthesis{Uno2024_PhD,
  author       = {Shinsuke Uno},
  title        = {{Wideband Multichroic Detector Architecture for Millimeter and Submillimeter Imaging Observations}},
  school       = {University of Tokyo},
  year         = {2024},
  address      = {Tokyo, Japan},
}

@Article{2024Inoue,
author={Inoue, Shuhei
and Chin, Kah Wuy
and Uno, Shinsuke
and Kohno, Kotaro
and Niwa, Yuka
and Naganuma, Toyo
and Yamamura, Ryosuke
and Watanabe, Kazuki
and Takekoshi, Tatsuya
and Oshima, Tai},
title={{A Design Method of an Ultra-Wideband and Easy-to-Array Magic-T: A 6-14 GHz Scaled Model for a mm/submm Camera}},
journal={Journal of Low Temperature Physics},
year={2024},
month={Jul},
day={01},
volume={216},
number={1},
pages={378-385},
issn={1573-7357},
doi={10.1007/s10909-024-03150-w},
url={https://doi.org/10.1007/s10909-024-03150-w}
}

@misc{2024Paine,
  author       = {Paine, Scott},
  title        = {{The am atmospheric model}},
  month        = sep,
  year         = 2024,
  publisher    = {Zenodo},
  version      = {14.0},
  doi          = {10.5281/zenodo.13748391},
  url          = {https://doi.org/10.5281/zenodo.13748391},
}

@article{2025Mroczkowski,
	author = {{Mroczkowski, Tony} and {Gallardo, Patricio A.} and {Timpe, Martin} and {Kiselev, Aleksej} and {Groh, Manuel} and {Kaercher, Hans} and {Reichert, Matthias} and {Cicone, Claudia} and {Puddu, Roberto} and {Dubois-dit-Bonclaude, Pierre} and {Bok, Daniel} and {Dahl, Erik} and {Macintosh, Mike} and {Dicker, Simon} and {Viole, Isabelle} and {Sartori, Sabrina} and {Valenzuela Venegas, Guillermo Andrés} and {Zeyringer, Marianne} and {Niemack, Michael} and {Poppi, Sergio} and {Olguin, Rodrigo} and {Hatziminaoglou, Evanthia} and {De Breuck, Carlos} and {Klaassen, Pamela} and {Montenegro-Montes, Francisco Miguel} and {Zimmerer, Thomas}},
	title = {{The conceptual design of the 50-meter Atacama Large Aperture Submillimeter Telescope (AtLAST)}},
	DOI= "10.1051/0004-6361/202449786",
	url= "https://doi.org/10.1051/0004-6361/202449786",
	journal = {A\&A},
	year = 2025,
	volume = 694,
	pages = "A142",
}

@unpublished{2025Uno, 
    author = {Shinsuke Uno and Kah Wuy Chin and Tai Oshima and Satoshi Ono and Takeshi Sakai and Kazuki Watanabe and Shuhei Inoue and Tatsuya Takekoshi and Kotaro Kohno}, 
    title = {{Design Method of Quasi-Lumped Element Bandpass Filters Using  Superconducting Coplanar Waveguide for Millimeter-Wave Multichroic Imaging}}, note = {submitted to \textit{IEEE Transactions on Applied Superconductivity}}, 
    year = {2025}
}

@unpublished{2026Naganuma, 
    author = {Toyo Naganuma and Shinsuke Uno and Shuhei Inoue and Kazuki Watanabe and Tatsuya Takekoshi and Takeshi Sakai and Tai Oshima}, 
    title = {{Broadband anti-reflection coating for sub-terahertz optics using dielectric multilayers}}, 
    note = {in press at \textit{Appl. Opt.}},
    year = {2026}
}

\vfill

\end{document}